\documentclass{du-journals}

\journal{Iranian Journal of Astronomy and Astrophysics}

\title{Noether Symmetry in f(T) Theory at the anisotropic universe}

\author[1]{A. Aghamohammadi}
\address[1]{Sanandaj Branch, Islamic Azad University, Sanandaj, Iran; email:a.aqamohamadi@gmail.com;\\a.aghamohamadi@iausdj.ac.ir}

\begin{document}

\begin{abstract}
As it is well known, symmetry plays a crucial role  in the theoretical physics. On other hand, the Noether symmetry is a useful procedure to select models motivated at a fundamental level, and  to discover  the exact solution to the given lagrangian. In this work, Noether symmetry in f(T) theory on a spatially homogeneous and anisotropic Bianchi type I universe is considered. We discuss the Lagrangian formalism of f(T ) theory in anisotropic universe.      The point-like Lagrangian is clearly constructed.The explicit form of f(T) theory and the corresponding exact solution are found by requirement of Noether symmetry and Noether charge. A power-law f(T), the same as the FRW universe, can satisfy the required Noether symmetry in the anisotropic universe with power- law scale factor. It is regarded that positive expansion is satisfied by a constrain between parameters.\\

\end{abstract}


\section{Introduction}
In this work, our aim is to study a Noether symmetry of scalar torsion gravity in anisotropic univers.\\
Recently, some astrophysical observations have shown that the Universe is undergoing an accelerated phase era. To justify this unexpected result, scientists have proposed some different models such as, scalar field models~\cite{1, 2, 3, 4} and modify theories of  gravity ~\cite{5, 6, 7, 8}. For the latter proposal, one can deal with telleparallel equivalent of general relativity~\cite{9,10, 11, 12}, in which the field equations are second order~\cite{13}. In addition, in this scenario the Levi-Civita connections are replaced by Weitzenb\"{o}ck connection where it has no curvature but only torsion~\cite{14}.\\

A Bianchi type I (BI) universe, being the straightforward generalization of the flat FRW universe, is of interest because it is one of the simplest models of a non-isotropic
universe exhibiting a homogeneity and spatial flatness.  In this case, unlike the FRW universe which has the same scale factor for three spatial directions, a BI universe has
a different scale factor for each direction. This fact introduces a non-isotropy to the system.
 The possible effects of anisotropy in the early universe have been  investigated with  BI  models from different points of view \cite{30,33,34,35}. Some people  \cite{Ro,Ko} have constructed cosmological models by using anisotropic fluid and BI universe. Recently, this model
has been studied in the presence of binary mixture of the perfect fluid and the DE  \cite{Ya}. Further, there are some exact solutions for BI models in f (T ) gravity
 \cite{Sh}
 \\

The outline of this work is as follows. In the next section, a brief review of the general formulation of the field equations in a BI metric and $f(T)$ garavity are discussed,  Sec. \ref{sec3}  is concerned  with Lagrangian formalism of f(T ) theory in anisotropic universe.    Sec. \ref{sec4}, is related to  Noether symmetry in f(T ) theory in anisotropic universe. We summarize our results in last section.

\section{General Framework}

The telleparallel theory of gravity is defined in the Weitzenb\"{o}ck's space-time, with torsion and zero local Riemann tensor, in which
we are working in a non-Riemannian manifold. The dynamics
of the metric were determined using the scalar torsion $T$ . The
fundamental quantity in teleparallel theory is the vierbein
(tetrad) basis $e^i_{\;\;\mu}$ . This basis is an orthogonal, coordinate free
basis defined by the following equation
\begin{eqnarray}
g_{\mu\nu}=\eta_{ij}e^{i}_{\;\;\mu}e^j_{\;\;\nu},
\end{eqnarray}
where $\eta_{ij}=diag[1,-1,-1,-1]$ and $e_{i}^{\;\;\mu}e^{i}_{\;\;\nu}=\delta^{\mu}_{\nu}$ or  $e_{i}^{\;\;\mu}e^{j}_{\;\;\mu}=\delta^{j}_{i}$. and the matrix $e^{a}_{\;\;\mu}$ are called tetrads that indicate the dynamic fields of the theory, where Latin $i, j$ are indices running over $0, 1, 2, 3$ for the tangent space of the manifold, and Greek
$\mu, \nu$ are the coordinate indices on the manifold, also running over $0, 1, 2, 3$.
In the framework of $f (T )$ theory, Lagrangian density is extended
from the torsion scalar $T$ to a general function $f (T )$, similar
to what happened in $f (R)$ theories. The action $S$ of modified
telleparallel gravity is given by \cite{Fe, Be}
\begin{eqnarray}\label{1bt}
 I=\int\mathrm{d}^4x|e|f(T)+L_m\Big],
  \end{eqnarray}
where for convenience, we use the units $2k^2=16\pi G=1,\,|e|=det(e^i_{\mu})=\sqrt{-g}$ and $e^i_{\mu}$ forms the tangent vector of the manifold, which is used as a dynamical object in telleparallel gravity, $L_M$ is the Lagrangian of matter.
The components of the tensor torsion and the contorsion are defined respectively as
\begin{eqnarray}
T^\rho_{\verb| |\mu\nu}\equiv e^{~\rho}_l
\left( \partial_\mu e^{l}_{~\nu} - \partial_\nu e^{l}_{~\mu} \right)\,,
\label{eq:2.2} \\
K^{\mu\nu}_{\verb|  |\rho}\equiv
-\frac{1}{2}
\left(T^{\mu\nu}_{\verb|  |\rho} - T^{\nu \mu}_{\verb|  |\rho} -
T_\rho^{\verb| |\mu\nu}\right)\,.
\label{2.3}
\end{eqnarray}
It was defined  a new tensor $S_\rho^{\verb| |\mu\nu}$ to obtain  the scalar equivalent to the curvature scalar of general relativity i.e. Ricci scalar, that is as
\begin{equation}
S_\rho^{\verb| |\mu\nu}\equiv \frac{1}{2}
\left(K^{\mu\nu}_{\verb|  |\rho}+\delta^\mu_\rho
T^{\alpha \nu}_{\verb|  |\alpha}-\delta^\nu_\rho
T^{\alpha \mu}_{\verb|  |\alpha}\right).
\label{2.5}
\end{equation}
Hence, the torsion scalar is defined by  the following contraction
\begin{equation}
T \equiv S_\rho^{\verb| |\mu\nu} T^\rho_{\verb| |\mu\nu}.
\label{7}
\end{equation}
By using the components (Eq.\ref{2.3},Eq.\ref{2.5}), the  torsion scaler (Eq. \ref{7}) gives
\begin{equation}\label{0bt}
T \equiv -6H^2+2\sigma^2.
\end{equation}
Bianchi cosmologies are spatially homogeneous but not necessarily isotropic. Here, we will consider BI cosmology. The metric of this model is given by
\begin{equation}\label{1}
ds^2=dt^{2}-A^{2}(t)dx^{2}-B^{2}(t)dy^{2}-C^{2}(t)dz^{2},
\end{equation}
where the metric functions, $A, B, C$, are merely  functions of time, $t$ and related to scale factor by $a=(ABC)^{\frac{1}{3}}$. In this work for  convenience, we assume $B=C=A^m$, where $m$ is a  constant. It is defined the shear tensor as describes the rate of distortion of the
matter flow, that in a comoving coordinate system,
from the metric   (Eq.\ref{1}), the components of the average Hubble parameter and the shear tensor
are given by \cite{36}
\begin{eqnarray}\label{10}
H&=&\frac{1}{3}(\frac{\dot{A}}{B}+\frac{\dot{B}}{B}+\frac{\dot{C}}{C}),\cr
\sigma^{2}&=&\frac{1}{2}\big[(\frac{\dot A}{A})^2+ (\frac{\dot B}{B})^2 + (\frac{\dot C}{C})^2 \big]-\frac{3}{2}H^2.
\end{eqnarray}


\section{Lagrangian Formalism of f(T ) Theory in Anisotropic Universe}
\label{sec3}
 In this section, we discuss the Lagrangian formalism of f(T ) theory in  anisotropic universe .In the study of Noether symmetry,It is clear that the point-like Lagrangian plays a crucial role. From the action f(T ) (Eq.\ref{1bt}), and following \cite{20, 21,22}, to derive the cosmological equations in the Bianchi I metric, (BIm ),
one can define a canonical Lagrangian ${\cal L} ={\cal L}(A, \dot{A}, T, \dot{T} )$, whereas $ Q = {a, T }$ is the configuration space,
and $ {\cal T}Q =[A, \dot{A}, T, \dot{T}]$ is the related tangent bundle on which L is defined. The  factor A(t) and the torsion scalar T (t) are taken as independent dynamical variables. One can use the method of Lagrange
multipliers to set T as a constraint of the dynamics  (Eq. \ref{7}). Selecting the suitable Lagrange multiplier
and integrating by parts, the Lagrangian ${\cal L}$ becomes canonical \cite{20, 21}
theory which is given by

\begin{eqnarray}\label{2bt}
 I=2\pi^2\int\mathrm{d}t ABC \big[ f(T)-\lambda(T+6H^2-2\sigma^2)-\frac{\rho_{m0}}{ABC}\big],
  \end{eqnarray}
where $\lambda$ is a Lagrange multiplier. The variation with respect to T of this action gives
\begin{eqnarray}\label{3bt}
\lambda=f_T.
  \end{eqnarray}
  So,  the action (\ref{2bt}) can be rewritten as
  \begin{eqnarray}\label{4bt}
 I=2\pi^2\int\mathrm{d}t ABC \big[ f(T)-f_T(T+6H^2-2\sigma^2)-\frac{\rho_{m0}}{ABC}\big],
  \end{eqnarray}
  and then the point-like Lagrangian reads (up to a constant factor $2\pi^2$)gives
   \begin{eqnarray}\label{4bt}
{\cal L}(A,\dot{A}, T, \dot{T})= A^{1+2m} \big[ f(T)-f_T(T+6H^2-2\sigma^2)\big]-\rho_{m0},
  \end{eqnarray}
  where using from assume $B=C,a^3=A^{1+2m}$.
 Writing (\ref{4bt}) with respect (\ref{10}) yield
 \begin{eqnarray}\label{5bt}
{\cal L}(A,\dot{A}, T, \dot{T})= A^{1+2m} \big[f-f_T T+2f_T(\frac{\dot{A}}{A})^2c_0 \big]-\rho_{m0},
  \end{eqnarray}
 where $c_0=1/3 (2m+1)-(1+2m)^2/3-m^2)$, and setting $m=1$ reduce equation (Eq.\ref{5bt}) to the same  form of lagrangian equation in isotropic univerce, i.e. $FRW$ metric \cite{22}, as well, the equation lagrangian form in the $FRW$ metric constrained  $c_0\neq 0,-2/7$.  As it is explicit for a dynamical system, the Euler-Lagrange equation is written
  \begin{eqnarray}\label{6bt}
  \frac{d}{dt}(\frac{\partial{\cal L}}{\partial{ \dot{ q_i}}})=\frac{\partial{\cal L}}{\partial q_i},
  \end{eqnarray}

  where $q_i$ are  $A, T$ in this case. Substituting Eq. (\ref{5bt}) into the Euler-Lagrange equation (Eq.\ref{6bt}), we get the following equations with respect $T, A$ respectively
 \begin{eqnarray}
  A^{1+2m}f_{TT}\big[-T+2(\frac{\dot{A}}{A})^2c_0 \big]=0,\label{7bt}\\
  4f_{TT}\dot{T}\dot{A}+4f_T\ddot{A}c_0+2f_T(\frac{\dot{A}}{A})^2(2m-1)c_0-(1+2m)(f-f_T T)=0.\label{8bt}
  \end{eqnarray}
  From Eq. (\ref{7bt}), it is easy to find that if $f_{TT}\neq 0$
  \begin{eqnarray}\label{9bt}
 T=2(\frac{\dot{A}}{A})^2c_0=-6H^2+2\sigma^2.
  \end{eqnarray}
 That, setting $m=1$  reduce equation (\ref{9bt}) to the same form of torsion scaler from $FRW$ metric, \cite{22} . In addition,  the relation (\ref{0bt}) is recovered. Mainly, this is the Euler constraint of the dynamics. Substituting
Eq. (\ref{9bt}) into Eq. (\ref{8bt}), we get
\begin{eqnarray}\label{10bt}
 8f_{TT}c_0^2\frac{\dot{A}}{A}\big[\frac{2\ddot{A}\dot{A}}{A^2}-\frac{2{\dot{A}}^3}{A^3} \big]+4f_Tc_0\frac{\ddot{A}}{A}+2f_T(\frac{\dot{A}}{A})^2c_0(2m-1)-(1+2m)(f-2f_Tc_0(\frac{\dot{A}}{A})^2)=0.
  \end{eqnarray}
This is  the modified Raychaudhuri equation, and by setting $m=1$ in the $c_0$ parameter, the above equation is reduced to the same equation in isotropic universe, i.e. FRW metric \cite{22}   . By the way,
it is explicit,  that the corresponding  Hamiltonian to Lagrangian ${\cal L}$ is given by
 \begin{eqnarray}\label{2f}
{\cal H}=\sum_i\frac{\partial {\cal L}}{\partial \dot q_i}\dot q_i-{\cal L}.
  \end{eqnarray}
Replacing (Eq.\ref{5bt}) into (Eq.\ref{2f}), one can rewrite the above Lagrangian density as follows
\begin{eqnarray}\label{3f}
{\cal H}=2f_T A^{2m-1}c_0 \dot A^2-A^{1+2m}(f-f_T T)+\rho_{m_0}.
 \end{eqnarray}
  Using the zero energy condition, ${\cal H}=0$, \cite{20,21,23}, we get
   \begin{eqnarray}\label{3f}
-2f_T A^{2m-1}c_0 \dot A^2+A^{1+2m}(f-f_T T)=\rho_{m_0},
 \end{eqnarray}
 where,  it is clear again that by taking $m=1$ in the $c_0$  parameter, the above equation end up to one in the modified Friedmann equation, i.e. the $f(T)$ gravity at $FRW$ metric. As a result, we have
found that the point-like Lagrangian obtained in (Eq.\ref{5bt}) can yield all the correct equations of motion in anisotropic universe, that taken $m=1$ in the $c_0$ parameter recovered  what is in isotropic universe, i.e. the $f(T)$ gravity equation in $FRW$ metric.
\section{Noether Symmetry in f(T ) Theory at Anisotropic Universe}
\label{sec4}
As mentioned, one can find the exact solution to the given lagrangian by using Noether symmetry theorem.
So in this section, we would like to investigate
Noether symmetry in f(T ) theory in anisotropic universe.
Following references  \cite{20,21,22} , the generator of Noether symmetry is a killing vector
 \begin{eqnarray}\label{4f}
X=\alpha\frac{\partial}{\partial \alpha}+\beta\frac{\partial}{\partial \beta}+\dot{\alpha}\frac{\partial}{\partial \dot{\alpha}}+\dot{\beta}\frac{\partial}{\partial \dot{\beta}},
 \end{eqnarray}
where $\alpha, \beta$, both are the function of the generalized coordinate of $A,T$. Requirement of Noether symmetry is that Lie differentiation  with respect $X$ to be zero. Hence we get
  \begin{eqnarray}\label{5f}
 L_X{\cal L}=\alpha\frac{\partial{\cal L}}{\partial \alpha}+\beta\frac{\partial{\cal L}}{\partial \beta}+\dot{\alpha}\frac{\partial{\cal L}}{\partial \dot{\alpha}}+\dot{\beta}\frac{\partial{\cal L}}{\partial \dot{\beta}}=0.
 \end{eqnarray}
Therefore, based on Noether symmetry theorem, there should be a  motion constant, so-called Noether charge \cite{20, 21}.
\begin{eqnarray}\label{6f}
 Q_0=\sum_i \alpha _i \frac{\partial {\cal L}}{\partial \dot{\alpha}_i }= \alpha\frac{\partial{\cal L}}{\partial \dot{A}}+\beta\frac{\partial{\cal L}}{\partial T}=\alpha(4f_T A^{2m-1}c_0\dot{A})=const,
 \end{eqnarray}
  where setting $m=1, c_0=-3$ in the above equation, recovered the same equation in \cite{22} to isotropic universe.
We know that $L_X {\cal L}=0$, meaning $\cal L$ is constant along the flow generated by $X$, i.e. (Eq.\ref{5f}\cite{23}. Therefore, evaluating (Eq.\ref{5f}) is a  second degree  function from $\dot A,  \dot T$, whose  coefficients are functions of $a$ and $T$ only. Hence, they have to be zero separately.
 So, replacing  (Eq.\ref{4bt}) into (Eq.\ref{5f}) and using the relations $\dot {\alpha}=\partial {\alpha}/\partial A \dot A+\partial {\alpha}/\partial T \dot T  $and $\dot {\beta}=\partial {\beta}/\partial A \dot A+
 \partial {\beta}/\partial T \dot T$ yield
 \begin{eqnarray}\label{7f}
 \alpha(1+2m)(f-f_T T)+2\alpha (2m-1)f_T(\frac{\dot A}{A})^2c_0-\beta Af_{TT}T+2\beta f_{TT}{\dot A}^2A^{-1}c_0+\cr
 4\frac{\partial \alpha}{\partial A}f_T{\dot A}^2A^{-1}c_0+4\frac{\partial \alpha}{\partial T}\dot T f_T\dot A A^{-1}c_0=0.
 \end{eqnarray}
 As mentioned above,  the coefficients $\dot A^2,\dot T \dot A$ should be zero, as a result, we get
 \begin{eqnarray}
 4\frac{\partial \alpha}{\partial T} f_T=0, \label{8f}\\
  2(2m-1)f_TA^{-2}\alpha +2\beta f_{TT}A^{-1}+4\frac{\partial \alpha}{\partial A} f_T A^{-1}=0,\label{9f}\\
 \alpha (1+2m)(f-f_T T)-\beta A f_{TT}T=0.\label{10f}
 \end{eqnarray}
 It is explicit that solutions of (Eqs. \ref{9f}, \ref{8f},\ref{10f})are given if  the explicit form of $\alpha, \beta$ are obtained, and if at least one of them is different from zero, then Noether symmetry exist\cite{21}.
From (Eq.\ref{8f}), it is clear that $\alpha$ is independent of $T$, so it is merely  a function of $A$. In addition,
 from (Eq.\ref{10f}), we get
 \begin{eqnarray}\label{1e}
  \alpha (1+2m)(f-f_T T)=\beta A f_{TT}T.
 \end{eqnarray}
By substituting (Eq.\ref{1e}) in to (Eq.\ref{9f}), we get

  \begin{eqnarray}\label{2e}
  2(2m-1)f_T T A^{-2}\alpha +2 \alpha(1+2m)(f-f_T T)A^{-2}+4f_TA^{-1}T\frac{\partial \alpha}{\partial A}=0.
 \end{eqnarray}
 By separation of variables, one can transform the above equation to two independent differential equations as follow
  \begin{eqnarray}\label{3e}
 1-\frac{A}{\alpha}\frac{\partial \alpha}{\partial A}=\frac{(1+2m)f}{2f_T T}.
 \end{eqnarray}
Since Right and left hand side are independent, hence, they must
be equal to a same constant, that fore convenience, we set $\frac{1+2m}{n}$. As a result, (Eq.\ref{3e}) is separated into two ordinary differential equations as
 \begin{eqnarray}
 1-\frac{A}{\alpha}\frac{\partial \alpha}{\partial A}=\frac{1+2m}{2n},\label{4e}\\
 \frac{(1+2m)f}{2f_T T}=\frac{1+2m}{2n}.\label{5e}
 \end{eqnarray}
It is readily obtained the solutions of these two above equation as follow
\begin{eqnarray}
f=\mu T^n, \label{6e}\\
\alpha=\alpha_0 A^{\frac{2n-1-2m}{2n}},\label{7e}
 \end{eqnarray}
 where, again the above equation is reduced into the same equation in isotropic universe by setting $m=1$, i.e. \cite{22}; hence it is the desired one and $\mu,\alpha_0 $ are integral constants. Substituting (Eqs.\ref{6e}, \ref{7e}) into (Eq.\ref{1e}) we get
 \begin{eqnarray}\label{8e}
\beta=-\frac{\alpha_0\mu(1+2m)}{n}A^{-\frac{1+2m}{2n}}T.
 \end{eqnarray}
 Up till now, we obtained the non-zero solution of $f(T), \alpha, \beta$. Therefore, Noether symmetry exists  in anisotropic universe on Bianchi  type I. Now, we try to obtain a solution of scale factor  for this $f(T)$ function. Hence substituting, the (Eqs.\ref{6e},\ref{7e},\ref{8e}) into (Eq.\ref{6f}) yields
 \begin{eqnarray}\label{9e}
\dot A A^{c_2}=(\frac{c_1}{c_0})^{\frac{1}{2}},
 \end{eqnarray}
where, $c_1=(Q_0/4\mu n c_0\alpha)^{1/n-1}$,$c_2=\frac{4m^2+2m-4mn-1}{4m(n-1)}$. It is readily obtained the solution of (\ref{9e}) as follow
 \begin{eqnarray}\label{10e}
A=(1+c_2)^{\frac{1}{1+c_2}}\big[(\frac{c_1}{c_0})^{\frac{1}{2}}t-c_3\big]^{\frac{1}{1+c_2}} ,
 \end{eqnarray}
where $c_3 $ is integral constant. From requirement $a{(t=0)}=0$, it is easy to see that the constant  $c_3$ is zero. As a result, $A\sim t^{\frac{1}{1+c_2}}$. Therefore,
as mentioned, relation between scale factor $a$  and component metric $A$ in anisotropic Bianchi type I with assuming, $B=C=A^m$  is $a^3=(ABC)=A^{1+2m}$, hence
\begin{eqnarray}\label{11e}
a\sim t^{\frac{1+2m}{3(1+c_2)}}.
 \end{eqnarray}
 Note that, requirement of positive expansion, $\ddot a>0$ requiring  a constrained between the $n$ and $m$, parameter as following
 \begin{eqnarray}\label{10e}
n-1<\frac{3m}{4+2m}.
 \end{eqnarray}
 It is clear, that one can readily obtained physical quantity corresponding to the exact solution $a$ and $f(T)$ namely, $H,\dot H, \sigma^2, $ and equation of state $\omega$, that in this work, it is not our scope.
\section{Discussion}
As it is well known, symmetry plays a crucial role  in the theoretical physics. On other hand, the Noether symmetry is a useful procedure to select models motivated at a fundamental level, and discover  the exact solution to the given lagrangian. In this work, Noether symmetry in f(T) theory on A spatially homogeneous and anisotropic Bianchi type I universe  have considered. We  have addressed  the Lagrangian formalism of f(T ) theory in anisotropic universe, and a Lagrangian form was obtained. The point-like Lagrangian was clearly constructed. The explicit form of f(T) theory and the corresponding exact solution were found by requirement of Noether symmetry and Noether charge. A power-law f(T),
have obtained in the anisotropic universe with power- law scale factor, that can satisfy the requirement of the Noether symmetry. It was regarded that positive expansion is satisfied. Our main conclusions can be summarized as follows
\begin{itemize}
\item A exact solution have been obtained to $f(T), a(t)$, that is reduced to those value in $FRW$ metric with selecting the $m$ parameter, equal one, i.e. $m=1$ and $c_0=-3$
   \item Requirement of  positive acceleration have obtained a constrained between $m$ and $n$  parameter, i.e. $n-1<\frac{3m}{4+2m}$
   \item To regain the equation lagrangian form in the $FRW$ metric, it was   required  $c_0\neq 0,-2/7$ .
   \item It was obtained a energy density matter form by the zero energy condition, ${\cal H}=0$,
\item We have  obtained  exact solution of scale factor $a$  by Noether charge condition.
\item At last,we have seen that the resulting f(T ) theory from Noether symmetry can be study in anisotropic universe that may be, plays an important role at early universe.
\end{itemize}

\end{document}